\documentclass[prb,twocolumn,showpacs,preprintnumbers,amsmath,amssymb,superscriptaddress]{revtex4}

\usepackage{graphicx}
\usepackage{dcolumn}
\usepackage{bm}
\usepackage{revsymb}
\usepackage{txfonts}

\begin{document}
\preprint{Accepted (PRB Brief Report)}

\title{Observation of zigzag and armchair edges of graphite\\using scanning tunneling microscopy and spectroscopy}

\author{Yousuke Kobayashi}
\email{ykobaya@chem.titech.ac.jp}
\author{Ken-ichi Fukui}
\author{Toshiaki Enoki}
\affiliation{Department of Chemistry, Tokyo Institute of Technology, 2-12-1, Ookayama, Meguro-ku, Tokyo 152-8551, Japan}

\author{Koichi Kusakabe}
\affiliation{Graduate School of Engineering Science, Osaka University, 1-3, Machikaneyama-cho, Toyonaka, Osaka 560-8531, Japan}

\author{Yutaka Kaburagi}
\affiliation{Department of Energy Science and Engineering, Musashi Institute of Technology, 1-28-1, Tamazutsumi, Setagaya-ku, Tokyo 158-8557, Japan}

\date{\today}

\begin{abstract}
The presence of structure-dependent edge states of graphite is revealed by both ambient- and ultra-high-vacuum- (UHV) scanning tunneling microscopy (STM) / 
scanning tunneling spectroscopy (STS) observations. On a hydrogenated zigzag (armchair) edge, bright spots are (are not) observed together with 
$(\sqrt{3}\times\sqrt{3})R30^{\circ}$ superlattice near the Fermi level ($V_{S} \sim -30$ mV for a peak of the local density of states (LDOS)) under UHV, 
demonstrating that a zigzag edge is responsible for the edge states, although there is no appreciable difference between as-prepared zigzag and armchair edges in air. 
Even in hydrogenated armchair edge, however, bright spots are observed at defect points, at which partial zigzag edges are created in the armchair edge.
\end{abstract}

\pacs{73.20.At  68.65.-k}
\maketitle
Finite-sized graphene has called attention for its peculiar electronic structure dependent on the dimensionality, size, and geometry. 
The interference effect on electronic wavefunctions can be dominant for two-dimensional (2D) structure, dependent on the sample size \cite{khari1,ykobaya}. Periodic 
contrast of the density of states (DOS) of carbon nanotubes, which consist of graphene rolled with chiral vectors, is due to electronic 
confinement effect appearing in the 1D electronic structure \cite{sglemay}. On top of those interference effects, edge-localized electronic states are more 
characteristic. Especially when those materials become smaller, the electronic structure drastically changes in case of nanometer-long carbon 
nanotubes or nanometer-wide graphene ribbons \cite{osusumu,knakada}. An analytical model for the distribution of edge-localized electrons of graphene and 
their density of states were proposed by Fujita, et al \cite{kkobaya,knakada,kwakaba}. According to the model, the non-bonding $\pi$-electrons at a zigzag edge 
can be delocalized toward the interior of the plane with a finite probability density, which is dependent on the wave number of edge states, 
The edge states make almost flat bands near the Fermi level in addition to $\pi$- and $\pi^{\ast }$-bands of graphene. 
Ferromagnetism can arise by an arrangement of the spins of non-bonding $\pi$-electrons at a zigzag edge of nanographene or graphene ribbon, 
on assumption of a model of the bipartite lattices \cite{khari2,kkusaka}. In contrast, those interesting characters are quite absent at an armchair edge. 
The peculiar LDOS due to the edge states near zigzag edge is supported by some experimental reports, for example, 
on disordered magnetism of activated carbon fibers or shouldered $I$-$V_{S}$ curve near the Fermi level of hydrogen-irradiated graphite 
step edge. However, the origin of atomic-structure-dependent LDOS, which is a key to solve the unconventional electronic structure and magnetism, 
remains unclear and still in mystery \cite{yshibaya,zklusek2}. Therefore direct observation of local electronic structure near an edge is the 
most important issue in clarifying the characters of edge states which relate to the experimental findings. Atomically resolved study 
about edges of graphite will be a strong support to previous theoretical/experimental papers and create a new bias to nanomaterials 
of graphite and its related materials. In the present Letter, we show STM images of zigzag and armchair edges of graphite near the Fermi 
level and $dI/dV_{S}$ curves of STS to observe the distribution of the edge-localized electrons and the edge states, 
accompanied with the theoretically calculated LDOS mapping to reproduce the experimental images.\\
\hspace*{10pt}All atomically resolved STM images in constant-height mode were taken at $V_{S} = 0.02$ V and $I = 0.7$ nA, using Pt-Ir tip by Nanoscope E 
(Digital Instruments, Co.) and UHV-STM (Unisoku, Co.) for observations in air and under UHV condition, respectively. Sample preparation 
of nanographite is given elsewhere \cite{amaffoun}. In the sample preparation process, pits can be also generated due to reaction of residue of oxygen 
with the HOPG substrate surfaces during the heat treatment \cite{zklusek1}. The samples were exposed to air after the sample preparation. As for the observation 
under UHV condition ($\sim 5\times 10^{-11}$ Torr), the prepared samples were heated at around 800 ${^\circ }$C to eliminate functional groups including oxygen as a form of 
CO \cite{kmorigu}, immediately followed by exposure to atomic hydrogen to terminate edges of graphite in a sample treatment chamber  (under UHV condition) connected to the STM 
observation chamber. A condition of the hydrogenation of the edges was set at the same condition with that for hydrogenation of Si(100) surface to make monohydride 
surface \cite{jjboland}. Adsorbed contaminants, which were introduced in the process of the sample preparation of nanographite on HOPG substrate or by the exposure to air, 
to the edges and a graphite surface can be removed by reactions with pure hydrogen during the hydrogenation process. By several repeats of the heat treatments and 
hydrogenation in the preparation chamber, the structure of the edges is arranged due to the removal of hydrocarbons from hydrogen-terminated edges \cite{tzecho1,tzecho2}.\\
\hspace*{10pt}The dispersion relation and two-dimensional LDOS mapping 
was calculated using the tight-binding approximation for $AB$-stacked double-layer graphene. The first layer represents 
top graphene layer with edges and the second layer represent the graphite substrate. The resonance integral and the 
overlap integral were parametrized using the Slater-Koster parameters \cite{jcslater} and were determind for 2$s$, 2$p$ 
orbital of carbon and 1$s$ orbital of hydrogen. Structural dependence of the parameters was determined following a previous 
literature for carbon \cite{dapapaco}. For carbon-hydrogen bonding, we fitted parameters of hydrogen to reproduce band 
structure of graphene strips with the zigzag edges obtained by the first-principles calculatoin with the local density 
approximation \cite{jpperdew,jyamauch}. Several percentage of displacement of carbon atoms near each edge was neglected in 
the H\"{u}ckel aprroximation. This makes a calculation tractable without harming essential features in DOS.\\
\hspace*{10pt}The obtained nanographene on highly oriented pyrolitic graphite (HOPG) substrate tends to have straight edges or polygonal structures. 
It is not difficult to find peculiar edge structures even in air. Figure 1 shows atomically resolved ambient-STM images of edges of nanographene 
whose diameter is about 50 nm for (a) and about 100 nm for (b). From the arrangement 
of honeycomb lattice or $(\sqrt{3}\times\sqrt{3})R30^{\circ}$ superlattice drawn in Fig. 1, the shapes of edges are zigzag and armchair types for (a) 
and (b), respectively. Bright spots were observed near both edges, in contrast to the theoretical prediction that those bright spots can be generated 
only by localized electrons at a zigzag edge \cite{osusumu,knakada}. Some irregular spots were observed near bright spots of the superlattice of an 
armchair edge in Fig. 1(b). They are situated at positions with smaller distances than the distance of nearest-neighbor $\beta$-atoms (0.246 nm). 
$dI/dV_{S}$ curves of STS could not be obtained on both edges with reproducibility.\\
\begin{figure}[h]
\includegraphics[width=7.0cm, clip]{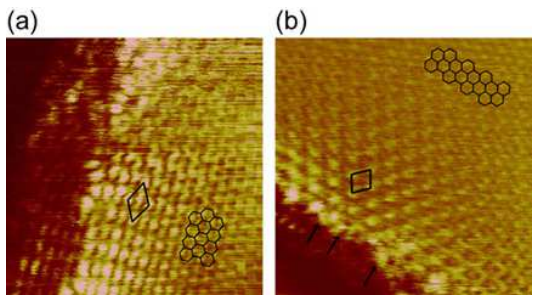}
\caption{\label{fig:edge1} Atomically resolved ambient-STM images ($5.6 \times 5.6$ nm$^{2}$) of (a) zigzag and (b) armchair edges of nanographene. 
For clarity of edge structures, models of honeycomb lattice and $(\sqrt{3}\times\sqrt{3})R30^{\circ}$ superlattice are drawn on the images and arrows are drawn to 
indicate irregular points at the armchair edge.}
\end{figure}
\hspace*{10pt}Figure 2(a) shows an atomically resolved UHV-STM image of hydrogenated step edge of HOPG. While no apparent contrast in spots were 
observed at center and bottom parts of the edge, bright spots were observed at top part of the edge. The top part of the edge is the 
zigzag type and the center and bottom parts correspond to the armchair type, judged from application of hexagonal lattice to the image near 
the edge. A typical $dI/dV_{S}$ curve near the bright spots is shown in Fig. 2(b). Peaks at about $-$0.03 and 0.2 eV were obtained 
accompanied with a little contribution of LDOS of $\pi$- and $\pi^{\ast }$-bands of graphite.\\
\begin{figure}[h]
\includegraphics[width=4.9cm, clip]{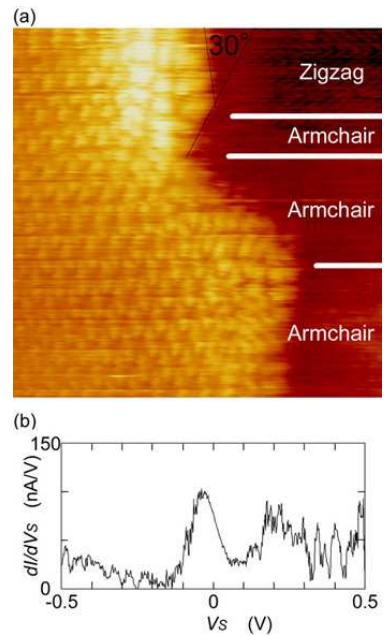}
\caption{\label{fig:edge2} (a) An atomically resolved UHV-STM image of zigzag and armchair edges ($9 \times  9$ nm$^{2}$). (b) Typical 
$dI/dV_{S}$ curve of STS data at a zigzag edge.}
\end{figure}
\begin{figure}[h]
\includegraphics[width=6.3cm, clip]{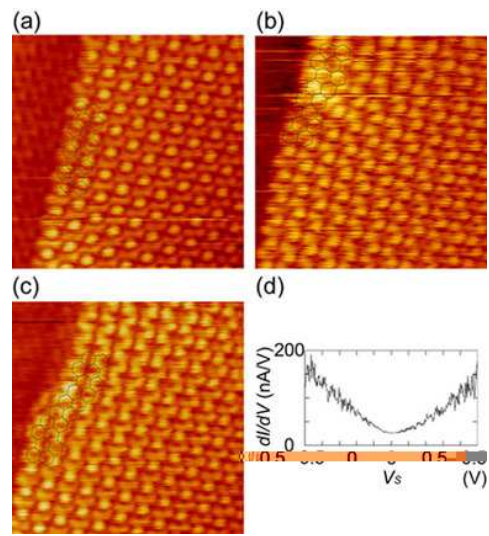}
\caption{\label{fig:edge3} Atomically resolved UHV-STM images ($5.6 \times 5.6$ nm$^{2}$) of (a) homogeneous armchair edge and (b, c) armchair edges 
with defect points. Two and four rows of armchair lines are added to the lower regions of the edges that start from the defect points 
in (b) and (c), respectively. For clarity of edge structures, models of honeycomb lattice are drawn on each image. (d) A $dI/dV_{S}$ curve of STS 
taken at the edge in (a).}
\end{figure}
\hspace*{10pt}Figures 3(a)-3(c) show atomically resolved UHV-STM images of parts of hydrogenated step edges of a pit, which is generated 
by reaction with residue of oxygen during the sample preparation and is about 40 nm by 70 nm in size. They are 
armchair edges of graphite, judged from the lattice information near the edges. The $dI/dV_{S}$ curve at the edge of Fig.3(a) is shown in Fig. 3(d). 
Only LDOS of $\pi$- and $\pi^{\ast }$-bands was observed and a contribution of peaks similar to that in Fig.2(b) was negligibly small.
In contrast to the image of the homogeneous armchair edge, that of defective armchair edges in Fig. 3(b) and 3(c) is obviously different.  
An array of bright spots, which shows decreasing LDOS toward the interior of the plane along a line with an angle of $60^{\circ }$ from the armchair edge, 
was observed at defect points in Fig. 3(b).  The defect consists of an increse of added two rows of carbon atoms to the armchair edge. However, 
such an array was not observed near the defect points in Fig. 3(c), where four rows of carbon atoms is added.\\
\hspace*{10pt}The discrepancy between the theoretical prediction and the ambient observations is due to random oxidation of the edges and 
adsorption of impurity atoms/molecules to the edges by exposing to air. The chemisorbed functional groups including oxygen atom change the LDOS 
at the edges. As another explanation for the discrepancy, one might think that the structure of carbon network at the observed armchair edge is destructed 
because the edge states can be observed dependent on edge structures near the Fermi level if as-prepared edges are hydrogen-terminated \cite{osusumu}. 
However, this is less suited for the description of Fig. 1(b), since we cannot specify the origin of the irregularity in the image as well as the bright spots damping 
toward the interior of the plane. To determine whether the edge states exist or not, we are required to specify the structure of edges under the UHV condition.\\
\hspace*{10pt}The LDOS dependence on the edge structures is clearly shown in Fig. 2(a), which is possible only for hydrogen-terminated samples 
under the UHV condition. The microscopy images prove that the edge states can be observed at a homogeneous zigzag edge and at a part of 
the armchair edge perturbed by an adjacent zigzag edge but they are not at armchair edges distant from other zigzag edges. The images, including 
$(\sqrt{3}\times\sqrt{3})R30^{\circ}$ superlattice, of homogeneous zigzag and armchair edges can be reproduced by calculated data in Ref.\onlinecite{plgiunta}. 
The STS data of Fig. 2(b) clearly verifies the presence of the edge states at the zigzag edge. In the figure, one peak at about $-$0.03 V corresponds to that of 
edge states and indicates that the flat band near the Fermi level in the theory in Ref.~\onlinecite{knakada,kwakaba}. Taking a rapid decay of 
the measured height from the edge to the interior of the plane into consideration, the flat band appears to be mainly around 
$k = \pi $ state, because the LDOS at $k = 2\pi /3$ state oscillates and does not decay \cite{knakada}. Origin of another peak at 0.2 V in Fig. 2(b) is attributed 
to charge transfer from a zigzag edge to physisorbed atoms/molecules. Taking the fact that hydrogenation process and the following STS observation 
are not completely free of impurity species and that the obtained STS data include little changes in relative position between the tip and the sample due to 
the thermal drift, this interpretation is reasonable \cite{zklusek2}. The image of armchair edges in Fig.2(a) is not homogeneous because the armchair edges are perturbed by the 
adjacent zigzag edge and corner points. Figure 3(a) clearly shows that a homogeneous hydrogenated armchair edge is created under the UHV condition. 
These facts demonstrate that the edge states are not observed on the homogeneous armchair edge, but the $dI/dV_{S}$ value of STS is not necessarily zero near the 
Fermi level due to the little charge transfer between the edge and the interior of the plane and due to week three-dimensionality of 
graphene layers.\\
\begin{figure}[h]
\includegraphics[width=4.6cm, clip]{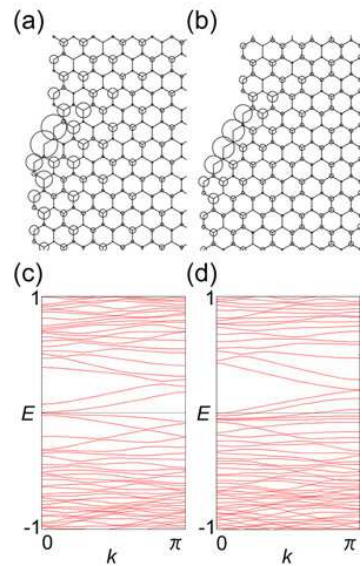}
\caption{\label{fig:edge4} (a, b) 2D mappings of the LDOS that reproduce the observed STM images using a tight-binding approximation 
for $AB$-stacked double-layer graphene; (a) and (b) correspond to the images of Figs. 3(b) and 3(c), respectively. The dimension of a 
circle on each lattice point denotes the relative value of the LDOS that is accumulated in the range of 50 meV near the Fermi level. 
The energy dispersions of (a) and (b) are shown in (c) and (d), respectively.}
\end{figure}
\hspace*{10pt}The origins of the bright points in Figs. 3(b)/3(c) are understood by looking at the LDOS of defect points of armchair edges. The shape of these defects is 
shown as increased rows of carbon atoms, as described by the honycomb lattice drawn in Fig 3(b)/3(c). We show 2D mappings of the LDOS of these defect structures 
in Fig. 4(a)/4(b). The tight-binding approximation for $AB$-stacked double-layer graphene is applied for the analysis, where edges are armchair type with two and four extra rows 
attached to the lower half of the armchair edge in Fig. 4(a) and 4(b), respectively. From these two figures, the center of distribution of the relatively large LDOS corresponds to the 
defect point of the increase of two/four rows of armchair edges, that is, partial zigzag edge embedded in an armchair edge. In Fig. 4(a), the mapping of the calculated LDOS shows 
a dispersed inclination of edge electrons and it roughly reproduces Fig. 3(b), despite that it fails to reproduce an angle of strong directivity of the bright points in Fig. 3(b). Contrasted to 
a case of an increase of two rows, Fig. 4(b) shows a localized inclination of edge electrons and it well reproduces isolated bright points, which are observed in Fig. 3(c), at the point of 
an increase of four rows. The figure also reproduces the $(\sqrt{3}\times\sqrt{3})R30^{\circ}$ superlattice near the point. The distribution of the LDOS, which depends on 
the shape of the defect points in Figs. 3(b) and 3(c), can be attributed to that around points at $k = 0$, which is shown in the crossing points in Figs. 4(c) and 4(d).\\
\hspace*{10pt}A possible explanation for the difference of the directivity is given by different edge structure at the defect points \cite{unpub}. Judged from the fabrication of 
the hydrogenated edge, it is possible that some extra carbon atom remains to bind to the defect points during the heat treatment process under the UHV condition. The carbon adatoms 
may change the electronic structure near the defect point. The array in Fig.3(b) is not observed in Fig.3(c). This may be because the edge states is energy-shifted or removed due to 
physisorption (or chemisorption) of atomic (or molecular) species. Another possible explanation may be given by imperfectness of hydrogen-terminated edge. Because hydrogen 
which is terminated at graphene edges cannot be detected near the Fermi level by STM, it is possible that a part of edges is hydrogen-missing. In the case, dangling-bond states can be 
generated. The Fermi level may be shifted downward, if dangling-bond states exist. An array of bright spots similar to that in Fig. 3(b) was observed by STM 
around single/a few atomic defects in graphene \cite{jamizes,pruffieu}. It is known that the direction of an array of bright spots around atomic defects in graphene depends on the 
underneath structure of carbon layers. Actually, an image of a defect point taken at an $\alpha $-site is different from that at a $\beta $-site \cite{kfkelly}. 
A similar effect is expected for a structure with a defective edge, and the directivity of the array in Fig. 3(b) may be understood as the site dependence of the defective edge.\\
\hspace*{10pt}In summary, edge states, which are dependent on edge structures, of graphene layers have been investigated by STM/STS. The edge states near the Fermi 
level are observed at a zigzag edge and defect points of an armchair edge, that is, at spin uncouple points. The edge states are not 
observed at a homogeneous armchair edge, although $(\sqrt{3}\times\sqrt{3})R30^{\circ}$ superlattice is observed dependent on the electronic states of the 
surroundings. Those experimental results reveal the dependence of the LDOS and the edge states of graphene layers on the edge structures.
 Other forms of edges of graphene can give a wide variety of LDOS near the edges near the Fermi level. To clarify the 
edge states of graphene, more experimental efforts are needed, that is, another periodic form of edges, another type of edge defect, 
or edges terminated with another chemical species, in the near future.\\
\hspace*{10pt}The authors are grateful to Prof. Hideo Aoki, for fruitful discussion. They also thank to Dr. A. Moore for his generous gift of HOPG 
substrate. The present work was supported by Grant-in-Aid for Scientific Research No. 15105005 from the Ministry of Education, Culture, 
Sports, Science, and Technology, Japan.
\bibliography{apssamp}
\end{document}